\documentclass[%
 reprint,
 amsmath,amssymb,
 aps,
]{revtex4-2}

\usepackage{graphicx}
\usepackage{dcolumn}
\usepackage{bm}
\usepackage{amsmath}
\usepackage{xcolor}
\usepackage{gensymb}

\usepackage{color,soul}

\begin{document}
\renewcommand{\figurename}{Fig.}
\preprint{APS/123-QED}

\title{Correlating microscopic viscoelasticity and structure of an aging colloidal gel using active microrheology and cryogenic scanning electron microscopy}

\author{Rajkumar Biswas}
 \email{rajkumar@rri.res.in}

\author{Vaibhav Raj Singh Parmar}%
 \email{vaibhav@rri.res.in}

\author{Anson G Thambi}%
 \email{anson@rrimail.rri.res.in}

\author{Ranjini Bandyopadhyay}%
 \email{Corresponding Author: Ranjini Bandyopadhyay;\\ Email: ranjini@rri.res.in}
\affiliation{Soft Condensed Matter Group, Raman Research Institute, C. V. Raman Avenue,
Sadashivanagar, Bangalore 560 080, INDIA}%

\date{\today}

\begin{abstract}

Optical tweezers (OTs) can detect pico-Newton range forces operating on a colloidal particle trapped in a medium and have been successfully utilized to investigate complex systems with internal structures. Laponite clay particles in an aqueous medium self-assemble to form microscopic networks over time as electrostatic interactions between the particles gradually evolve in a physical aging process. We investigate the forced movements of an optically trapped micron-sized colloidal probe particle, suspended in an aging Laponite suspension, as the underlying Laponite microstructures gradually develop. Our OT-based oscillatory active microrheology experiments allow us to investigate the mechanical responses of the evolving microstructures in aging aqueous clay suspensions of concentrations ranging from 2.5\% w/v to 3.0\% w/v and at several aging times between 90 and 150 minutes. We repeat such oscillatory measurements for a range of colloidal probe particle diameters and investigate the effect of probe size on the microrheology of the aging suspensions. Using cryogenic field emission scanning electron microscopy (cryo-FESEM), we examine the average pore areas of the Laponite suspension microstructures for various sample concentrations and aging times. By combining our OT and cryo-FESEM data, we report here for the first time to the best of our knowledge, an inverse correlation between the crossover modulus and the average pore diameter of the aging suspension microstructures for the different suspension concentrations and probe particle sizes studied here.

\end{abstract}

\maketitle

\section{Intoduction}

As a result of the pioneering work of Arthur Ashkin and coworkers~\cite{Ashkin:86}, optical tweezers (OTs) have become an invaluable tool for studying various processes in physical and biological systems~\cite{kinesin,dnaunzipping,cellshorting,anderson2018,neckernuss2015active,biosamples}. An OT utilizes the forces exerted by a focused laser beam to confine a micron-sized dielectric particle suspended in a medium~\cite{otreview}. It can operate as a non-invasive force transducer, which can accurately produce and measure pico-Newton (pN) range forces and nanometer (nm) range displacements with high temporal resolution (down to a few $\mu$s)~\cite{jones2015optical,Gieseler:21}. Their utility as a single molecule manipulation technique makes OTs an important resource for investigating various processes, such as in quantifying the kinetics of kinesin motors walking on microtubules~\cite{kinesin}, in unzipping of DNA~\cite{dnaunzipping}, in cell sorting~\cite{cellshorting}, and in measuring the mechanical responses of soft materials, such as colloidal suspensions~\cite{Jop_2009,PhysRevLett.98.108302}, polymer networks~\cite{neckernuss2015active,anderson2018} and bio-samples~\cite{biosamples}.

The study of the flow and deformation of materials and the determination of their mechanical properties is known as rheology \cite{macosko1994rheology}. Conventional viscometers and rheometers can measure mechanical properties of a material even with sample volumes less than a milliliter~\cite{rheologyreview}. Passive microrheology requires a much smaller sample volume and can measure linear viscoelastic responses in a broad frequency domain by tracking the thermal motion of a probe particle suspended in a medium~\cite{mason1995optical}. This technique uses the generalized Stokes-Einstein relation to connect the mean square displacement (MSD) of the probe particle with the viscoelastic properties of the medium~\cite{waigh2005microrheology,puertas2014microrheology}. Video particle tracking microrheology~\cite{CROCKER1996298}, dynamic light scattering~\cite{dlsmicrorheology}, magnetic tweezers~\cite{magnetictweezermicrorheology} and optical tweezers~\cite{Tassieri2010measuring} have been used successfully to study the viscoelastic properties of a range of materials. Passive microrheology provides useful information on local micro-structures of the material and has been used to probe spatial heterogeneities and non-ergodicity in the suspending medium ~\cite{heterogeneity}.

The measured microrheological properties of viscoelastic materials depend on the size of the trapped probe particle~\cite{probeize,VALENTINE20044004,PhysRevE.83.041902}. The surface chemistry of the probe particle and its interaction with the medium also influence the measurement~\cite{VALENTINE20044004,PhysRevE.83.041902}. Valentine \textit{et al.}~\cite{VALENTINE20044004} showed that in the absence of interactions between the probe and viscoelastic medium, a probe size larger than the characteristic lengthscale relevant to the viscoelastic medium is necessary to interpret the results accurately. Interestingly, the authors also noted that elasticity of the viscoelastic matrix, which arises due to the presence of suspension microstructures, can be measured using a relatively smaller probe particle in the presence of significant probe-medium interactions. Owing to the smaller sizes of the probe particles in this last case, the measurements are not sensitive to the spatial heterogeneity present in the viscoelastic medium. 

He \textit{et al.} ~\cite{PhysRevE.83.041902} showed that surface adsorption on a probe particle embedded in a network of actin filaments results in an enhanced estimate for the stiffness of the biopolymer network. An earlier work had proposed a scheme to study colloidal forces, structure, dynamics and rheology of attractive colloidal gels by combining microscopy, OT and rheology \cite{hsiao2014model}. Passive microrheology becomes less effective with highly viscous or gel-like samples due to reduced motion of the probe particle, and a simultaneous reduction in the signal-to-noise ratio~\cite{whyactive}. In such cases, active microrheology can be used, wherein an external force is applied to the probe particle and its resultant motion is measured to determine the viscoelastic properties of the surrounding matrix. Optical tweezers~\cite{Tassieri2010measuring}, magnetic tweezers~\cite{PhysRevLett.77.4470} and atomic force microscopy~\cite{rigato2017high} are the most commonly used active microrheology techniques. Active microrheology with optical tweezers can be implemented using two methods, $viz.$, oscillatory measurements and creep measurements~\cite{anderson2018}. In oscillatory measurements, the sample stage is moved in an oscillatory fashion, and the resultant force experienced by the trapped probe particle is computed. The amplitudes of the two signals and the phase difference between them are analyzed to determine the viscoelastic properties of the medium~\cite{anderson2018}. On the other hand, in creep measurements, a constant stress is applied to the medium by displacing the probe particle, following which the response of the material is measured. The oscillatory active microrheology technique is more closely related to conventional bulk rheology as its output is in the form of the frequency-dependent storage and loss moduli, G$^\prime$ and G$^{\prime\prime}$ respectively, of the sample~\cite{anderson2018}.

  Laponite$^\text{\textregistered}$ clay is used as a rheological modifier in many industrial products such as coatings, personal care products and paints. Laponite is a synthetic clay with disk-shaped particles of diameters in the range 25 - 30 nm and thickness $\approx$ 1 nm. In powder form, Laponite particles are arranged together in one-dimensional stacks called tactoids, with Na$^+$ ions residing in the inter-gallery spaces ~\cite{van1977introduction}. When Laponite powder is mixed with water, Na$^+$ ions from the inter-gallery spaces diffuse to the bulk to reduce osmotic pressure gradients, and the faces of the Laponite particles acquire negative charges. Laponite particles in the tactoids therefore repel each other, resulting in tactoid swelling and exfoliation. The particles in the aqueous suspension acquire positive charges at the rims due to hydration of the constituent magnesia groups~\cite{TAWARI200154}.  Due to the exfoliation of Laponite platelets from the tactoids and the simultaneous leaching of Na$^+$ ions from the particle faces, the electrical double layer surrounding each Laponite particle shows a gradual and continuous evolution over timescales upto a few hours~\cite{ali2015evaluation, shahin}. An increase in the number of counter-ions participating in the electric double layer results in the gradual reduction of inter-particle electrostatic repulsive forces. The positive edge of a Laponite particle interacts with the negatively charged face of another Laponite particle and initiates the buildup of suspension microstructures $via$ overlapping coins (OC) and house of cards (HoC) configurations~\cite{C2SM25731A}. As time passes, Laponite particles self-assemble to form system-wide fragile microstructures in a physical aging process that enhances the elasticity of the aqueous Laponite suspension with time~\cite{saha2014investigation,saha2015dynamic,misra}. Laponite suspensions between the concentrations $1.75\%$ w/v and $3.5\%$ w/v form soft glasses and exhibit physical aging behavior \cite{bandyopadhyay2006slow, mourchid1995phase}. An aging time $t_w$, the time elapsed after preparation of the Laponite suspension, is used to parameterize the physical aging of the suspension. The spontaneously evolving structures constituted by the Laponite particles break or destructure when shear is applied to the suspension. After the removal of external shear forces, the networks again reassemble or restructure with time.

  Aqueous suspensions of Laponite have been widely used as model systems to probe the properties of non-ergodic states such as gels and glasses~\cite{Laporeview}. We have recently shown that  falling ball viscometry can be used successfully to quantify the shear thinning property of a freshly prepared Laponite sample in terms of the destructuring rate of the suspension~\cite{RajkumarFB}. The time-dependent aging behaviour of soft colloidal glasses and their dependence on the physicochemical properties of the suspension have been studied extensively using macroscopic rheology and passive microrheology~\cite{PhysRevLett.93.160603,PhysRevLett.98.108302, Jop_2009,probeize,PhysRevE.78.021405}. There are a few reports on the passive microrheology of soft glassy clay suspensions that validate the fluctuation-dissipation theorem in these aging non-equilibrium systems~\cite{PhysRevLett.93.160603,PhysRevLett.98.108302, Jop_2009}. Another passive microrheology study of Laponite suspensions reported that microscopic mechanical properties approach those extracted from bulk rheology measurements as the size of the embedded probe is increased~\cite{probeize}, with the viscoelastic liquid to soft solid transition occurring earlier on the bulk scale than on the microscopic scale~\cite{PhysRevE.78.021405}. However, to the best of our knowledge, oscillatory active microrheology has never been performed to study the length-scale dependent mechanical response of aging aqueous Laponite clay suspensions.

In this report, we employ oscillatory active microrheology to investigate the mechanical properties of aqueous Laponite suspensions at various aging times $t_w$ and concentrations $C_L$. We apply sinusoidal oscillations to the sample cell and measure the force experienced by an optically trapped probe particle that is physically attached to the suspension microstructure network. The piezo-controlled stage is oscillated over a range of frequencies to determine the frequency-dependence of the elastic (G$^\prime$) and viscous (G$^{\prime\prime}$) moduli of Laponite suspensions of different $C_L$ and at various $t_w$. We observe that G$^{\prime\prime}$ dominates over G$^{\prime}$ at smaller stage oscillation frequencies, $\omega$, while G$^\prime$ exceeds G$^{\prime\prime}$ when $\omega$ is increased beyond a crossover frequency, $\omega_{co}$. The characteristic relaxation time $\tau$ of the suspension varies as the inverse of the crossover frequency, $\tau = 2\pi/\omega_{co}$. The crossover modulus $G_{co}$ is the modulus value where G$^{\prime}$ = G$^{\prime\prime}$. We report a systematic increase in $G_{co}$ as the concentration of the suspension, $C_L$, is increased, thereby signaling the onset of a jamming transition.  Furthermore, at a fixed $C_L$, we see that $G_{co}$ increases with aging time $t_w$ for the range of aging times explored here. We also investigate the length-scale dependence of $G_{co}$ by varying the size of the probe particle. Cryogenic field emission scanning electron microscopy (cryo-FESEM) images of Laponite suspensions at various $t_w$ and $C_L$ reveal that the average pore diameters $<D_p>$ of the underlying fragile suspension microstructures do not change significantly with $t_w$ but decrease with increasing $C_L$, for the ranges of $C_L$ and $t_w$ explored here. Finally, we show using different Laponite concentrations and probe diameters that the crossover modulus $G_{co}$ of the suspension is inversely related to the average diameter of the microscopic pores of the suspension structures that evolve spontaneously and gradually due to the physical aging process.

\section{Material and Methods}

\subsection{Preparation of Aging Laponite Clay Suspensions:}

Laponite$^\text{\textregistered}$ (BYK Additives Inc.) is a synthetic clay comprising disk-shaped particles (25-30 nm diameter and 1 nm thick). Laponite XLG has been used for all the experiments reported in this work. Hygroscopic Laponite powder is baked for 16-20 hrs at 120$^{\circ}$C to remove moisture. Predetermined amounts of the dried powder is weighed and added to 50 ml Milli-Q water (Millipore Corp., Resistivity = 18.2 M$\Omega$.cm) with continuous stirring to make a suspension of the desired concentration. After 40 minutes of mixing, 15 ml of the freshly prepared suspension is filtered using a syringe filter (pore size 0.45 $\mathrm{\mu}$m, Sigma Aldrich) into a small glass vial. A dilute aqueous suspension of polystyrene (PS) beads is next added to the freshly prepared suspension and stirred for 10 minutes. The resulting Laponite suspension containing suspended polystyrene beads is loaded into a sample cell and the open end is sealed using UV glue. In our experiment, zero waiting time ($t_w$ = 0) coincides with the time at which the stirring of the suspension is stopped. The sample cell is then kept on the piezo stage of the optical tweezer to initiate the trapping procedure.

\begin{figure}
\includegraphics[scale=.15]{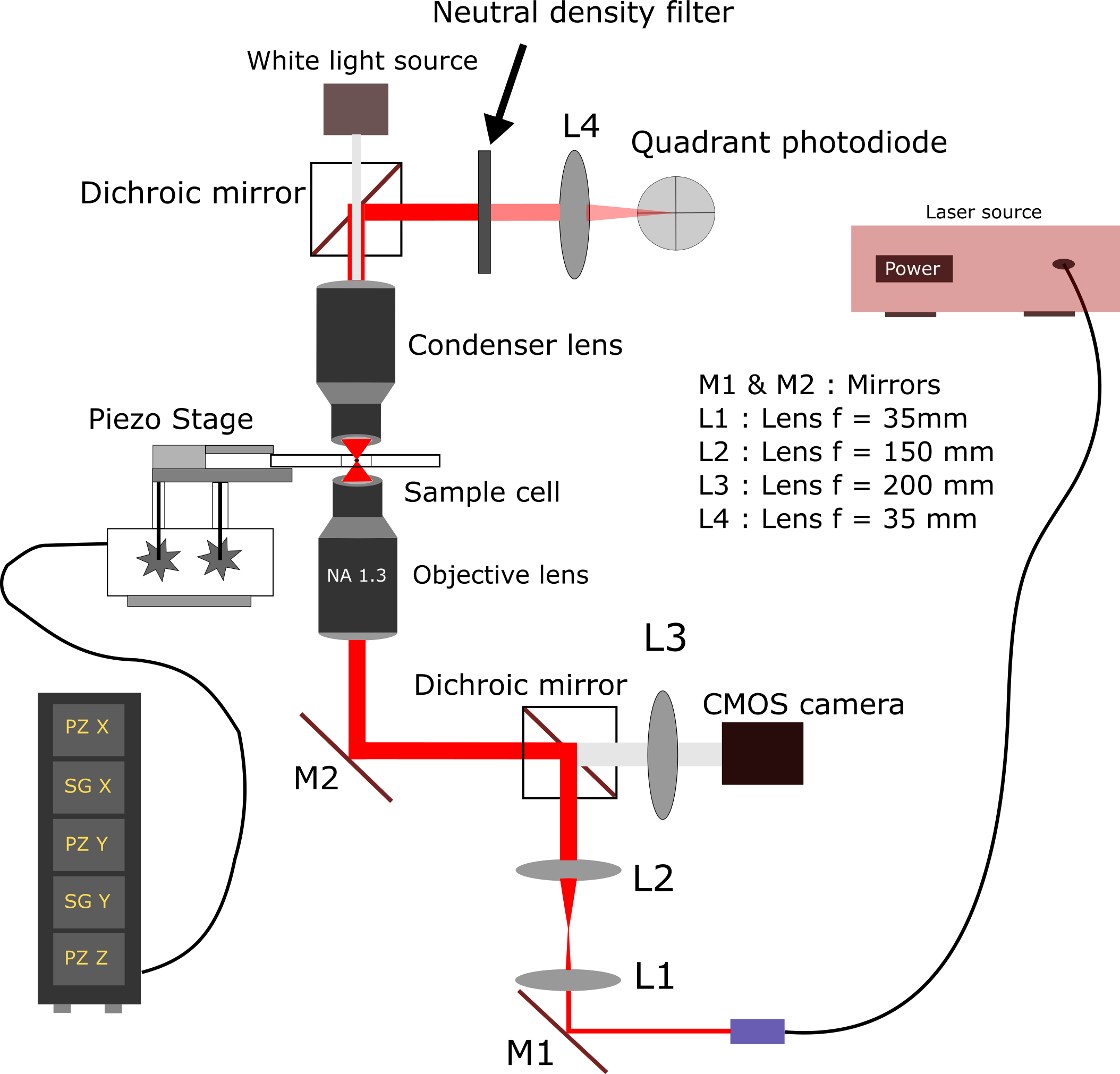} 
\caption{\label{fig:schematic}Schematic diagram of the optical tweezer set-up. The infrared laser beam is indicated in red.}
\end{figure}

\subsection{Optical Tweezer}

We use a Thorlabs optical tweezer (OT) module for our experiments. The schematic illustration of the OT setup is shown in Fig.~\ref{fig:schematic}. A continuous wave solid-state fiber laser (YLR-5-1064-LP, IPG Photonics USA) generates a Gaussian beam profile of wavelength 1064 nm. The beam is expanded by lenses L1 and L2 and deflected using dichroic and plane mirrors. The resultant beam is incident on a high numerical aperture (NA = 1.3) objective lens (oil immersion) of magnification 100X which focuses the beam to a diffraction-limited spot inside a sample cell. The highly focused beam can trap an object in three dimensions. The light scattered by the trapped particle is collected by a condenser lens (Nikon 10x), diverted using another dichroic mirror-lens assembly, and imaged on a position sensing quadrant photodiode (QPD - PDQ80A Thorlabs, Fig.~\ref{fig:schematic}) to measure the nanometer scale displacements of the trapped probe particle and the force exerted on it. The voltage signals from the QPD are recorded at a sampling frequency of 1000 Hz using a data acquisition card (National Instruments, USA) and custom software (LabView 2021). A sample cell containing the Laponite suspension with PS beads is mounted on the piezoelectric sample stage (TBB1515/M, Thorlabs), and oscillated at the desired frequency using a 3-axis XYZ translational stage (NanoMax-TS, Thorlabs), three piezoelectric actuators (TPZ001, Thorlabs) and two strain gauges (TSG001, Thorlabs). The sample cell is constructed using one glass slide and one \#1 coverslip (thickness $\approx$ 150 $\mu$m, Blue Star, India) and two \#0 coverslips (thickness $\approx$ 100 $\mu$m) as spacers, as shown in Supplementary Fig. S1. All the experiments were performed at 22$^{\circ}$C. The calibration of the piezoelectric stage and QPD are discussed in Supplementary Information Sections ST2 and ST3 respectively. The trap stiffness, used for calculating the force experienced by the probe particle, is evaluated using the power spectral density (PSD) method. The stiffness calculation of the optical trap in an aqueous medium for various laser powers is discussed in Supplementary Information Section ST4.

\begin{figure*}[ht]
\includegraphics[scale=.4]{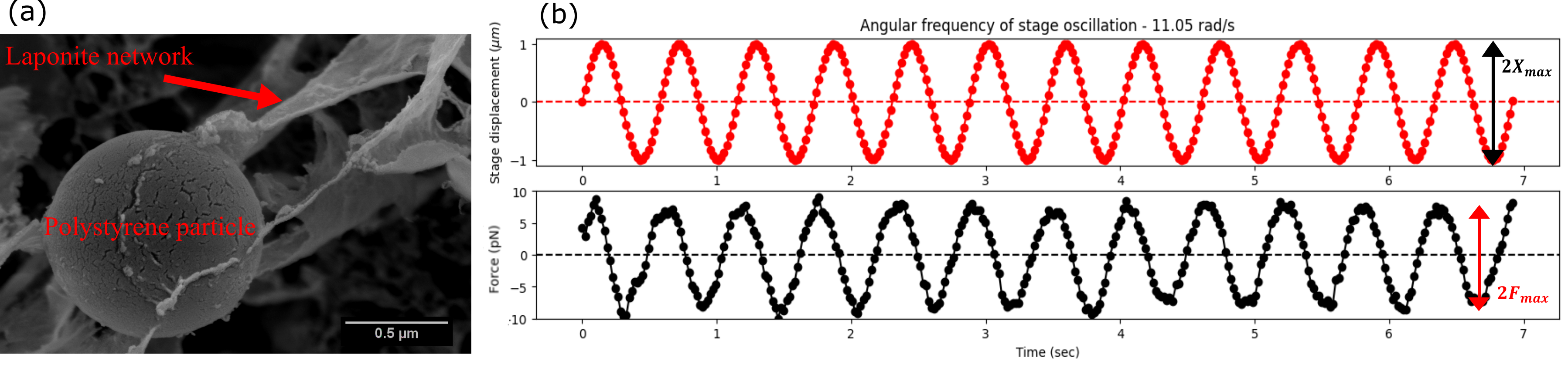} 
\caption{\label{oscillationAndCryo} \textbf{(a)} Cryo-FESEM image of a polystyrene particle attached to the Laponite suspension microstructure \textbf{(b)} Top: Sinusoidal displacement of the sample stage vs. time; Bottom: Force experienced by the trapped probe particle.}
\end{figure*}

\subsection{Cryogenic Field Emission Scanning Electron Microscopy (Cryo-FESEM)}

The evolving microstructures of the aging Laponite suspension, prepared at various aging times $t_w$ and Laponite concentrations $C_L$ using the protocol detailed in IIA, were imaged using a field emission scanning electron microscope from Carl Zeiss with an electron beam strength of 5 kV. The freshly prepared Laponite suspension was loaded in a capillary tube (Capillary Tube Supplies Ltd, UK) of diameter 1 mm using capillary flow. The capillary tube was sealed and kept undisturbed at room temperature for a predetermined aging time. The tubes loaded with the sample were then vitrified quickly in liquid nitrogen at -207$^{\circ}$C and transferred into a vacuum chamber at -150$^{\circ}$C (PP3000T Quorum technology). The sample was cryo-fractured using an in-built knife, and sublimated at a temperature of -90$^{\circ}$C for 20 minutes to remove the water content at the upper surface. Finally, a layer of platinum was coated on the sample surface for enhanced contrast during scanning electron microscopy. The surface images of the Laponite suspension were produced by capturing back-scattered secondary electrons. Laponite networks were detected from the cryo-FESEM images using ImageJ software (Wayne Rasband, NIH, US) and pore sizes were calculated. The mean value of pore diameter was measured by averaging over 10-50 different pores.

\begin{figure}[h]
\includegraphics[scale=0.7]{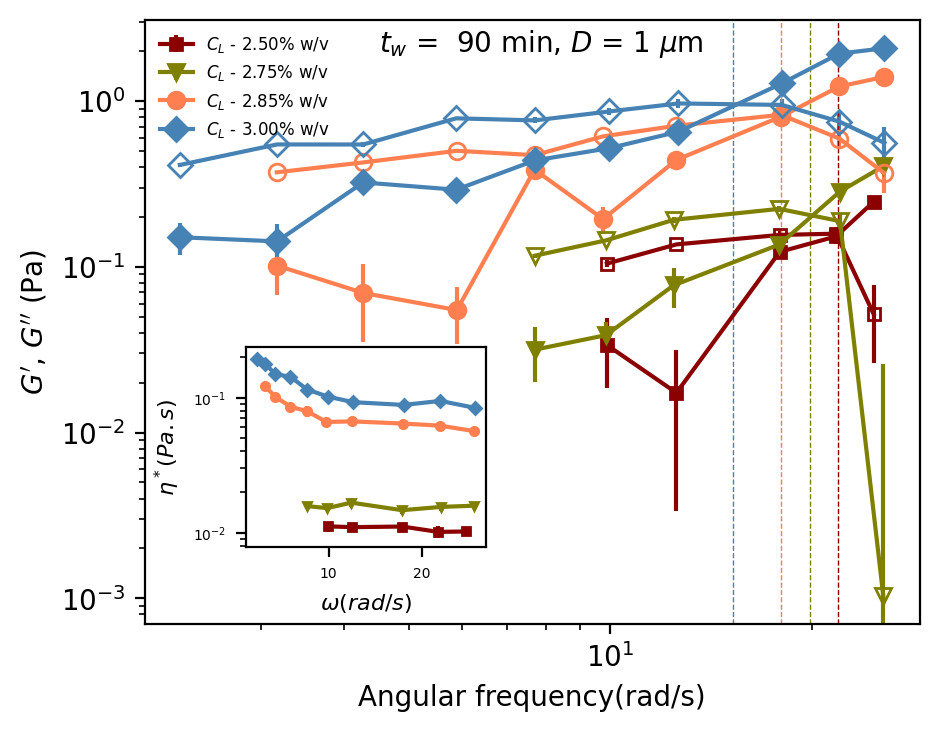}
\caption{\label{Concentration} Elastic moduli $G'$ (solid symbols) and viscous moduli $G''$ (hollow symbols) obtained in microrheological measurements using a probe particle of diameter $D=1$ $\mu$m and plotted $vs.$ angular frequencies of the sample stage for different Laponite suspension concentrations $C_L$ at an aging time $t_w$ = 90 min. The vertical dashed lines represent the crossover frequencies where $G^{\prime} = G^{\prime\prime}$ for data acquired at different $C_L$. In the inset, the frequency-dependent complex viscosities $\eta^* = \sqrt{G'^2 + G''^2}/\omega$ are plotted for the same Laponite concentrations. }
\end{figure}

\begin{figure*}[ht]
\includegraphics[scale=0.54]{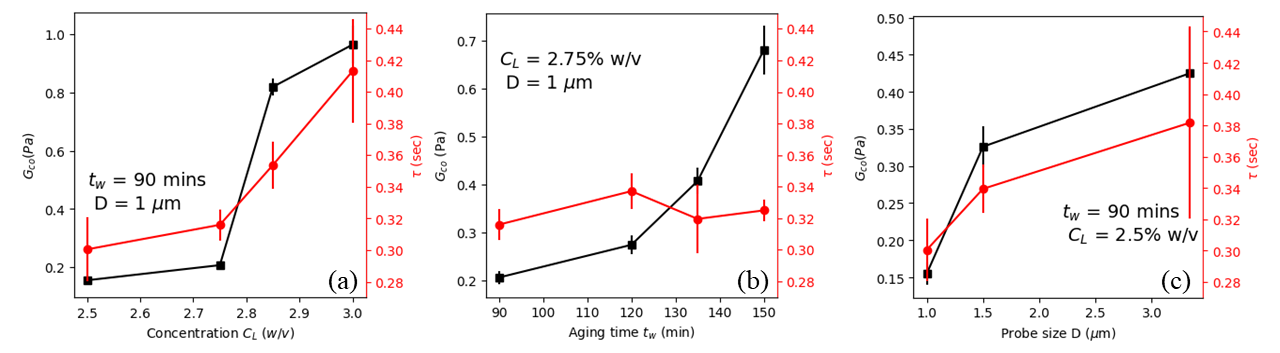}
\caption{\label{G_co} Crossover modulus $G_{co}$ (black) and  relaxation time $\tau$ (red) of the underlying Laponite suspensions are plotted for different (a) Laponite suspension concentrations $C_L$ (b) Laponite suspension aging times $t_w$ and (c) trapped probe particle sizes $D$.
}
\end{figure*} 

\section{ Results and Discussion}

An aqueous Laponite suspension shows physical aging~\cite{aging,ali2015evaluation}, with the suspension viscoelasticity spontaneously increasing due to the gradual formation of medium-spanning self-assembled microstructures~\cite{aging2}. The microstructures can rupture or destructure under shear forces. When the shear is removed, these structures heal or restructure spontaneously. For our OT-based active microrheology experiments, we study the dynamics of polystyrene particles trapped in the Laponite medium. Information about interaction between the probe and Laponite suspension microstructures is necessary for interpretation of microrheology measurements~\cite{VALENTINE20044004,PhysRevE.83.041902}. We perform isothermal calorimetry and cryogenic scanning electron microscopy images to visualize the extent of probe-Laponite attachment. Isothermal calorimetry indicates an exothermic reaction ($\approx$ 350-560 mJ/g) which we attribute to the attachment of PS to Laponite network strands. This energy release is much higher than that estimated between Laponite particles due to attractive interactions ($\approx$ 200 mJ/g \cite{probeize}) and confirms probe particle attachment to Laponite networks. Probe-Laponite attachment is further supported by the cryo-FESEM images in Fig.~\ref{oscillationAndCryo}(a) and Supplementary Fig. S5 and can be explained by considering attachment of the positive rim of a Laponite particle to the negatively-charged surface of a probe particle as schematically illustrated in Supplementary Fig. S6. Since the attachment of the probe particle with suspension microstructure obstructs its thermal motion, we choose to implement active microrheology to probe the Laponite system by applying an external force larger than the thermally induced fluctuations of the probe particle.

We perform active oscillatory microrheology~\cite{ziemann1994local} using optical tweezers (OT) to extract the elastic and viscous moduli, G$^\prime$ and G$^{\prime\prime}$ respectively, of an aging Laponite clay suspension in the presence of a very low volume fraction ($10^{-6}$ \% v/v) of probe particles. The combination of high-accuracy position sensing of the probe using a quadrant photodiode (QPD) and highly sensitive measurements of displacements of the sample stage using piezoelectric actuators allow us to acquire data with micrometer precision. While the probe particle is optically trapped in the Laponite suspension, we oscillate the sample cell sinusoidally (Fig.~\ref{oscillationAndCryo}(b) top panel) along the x-axis (as shown in Supplementary Figure S1) with a maximum amplitude of 1 $\mathrm{\mu}$m for a range of stage oscillation frequencies $\omega$. The oscillatory movement of the sample stage results in a net oscillatory force experienced by the optically trapped probe, which is calculated by measuring its nanometer-scale oscillatory displacement. The displacement of the probe from the trap center, caused by the sinusoidal motion of the piezoelectric stage, is converted to the corresponding spring force using the trap stiffness calculated as discussed in Supplementary Fig. S4. The microscopic elastic and viscous moduli, $G'$ and $G''$ respectively, of the underlying suspension are then calculated using the relations ~\cite{ziemann1994local}: 
\begin{equation} \label{Gprime}
G^\prime =  \frac{F_{max}}{6 \pi R X_{max}} \cos(\Delta \phi ),
\end{equation}
\begin{equation} \label{GDoubleprime}
G^{\prime \prime}=  \frac{F_{max}}{6 \pi R X_{max}} \sin(\Delta \phi )
\end{equation}
where $F_{max}$ is the maximum force acting on the trapped particle, $R$ is the radius of the trapped particle, $X_{max}$ is the maximum amplitude of stage oscillation, and $\Delta \phi$ is the phase difference between the displacement of the sample stage and the force experienced by the trapped particle measured in radians as shown in Fig.~\ref{oscillationAndCryo}(b). 

We oscillate the sample stage at various frequencies ranging from $\approx$ 3 rad/s to 24 rad/s, with the maximum oscillation frequency limited by the speed of data transfer during a given time interval. Simultaneously, the minimum oscillation frequency is set by the relative contributions of the viscous drag and  thermal forces acting on the trapped probe particle. The force experienced by the probe at a very small stage oscillation frequency results in a very low signal to noise ratio as shown in Supplementary Fig. S7. We ensure an accurate measurement of the oscillatory force response of the bead by increasing the stage frequency above a certain threshold value that depends on Laponite concentration $C_L$ and aging time $t_w$. We oscillate the stage with a fixed amplitude and measure the oscillatory drag force on the probe and phase difference between stage displacement and measured drag force at a fixed frequency. We estimate the elastic and viscous moduli, $G^{\prime}$ and $G^{\prime \prime}$, of the Laponite suspension by repeating stage oscillations at each frequency for 12 cycles to ensure adequate statistics. Error bars are calculated according to the procedure described in Supplementary Information Section ST6. Two or three independent repetitions have been performed for selected samples to confirm data reproducibility. We note here that since all the measurements were performed for Laponite ages $t_w$ ranging from 90 to 150 minutes, the evolution of the viscoelastic moduli during the data acquisition interval of 5 minutes is expected to be minimal. The above range of aging times is chosen to ensure that the viscosity contribution of the underlying medium causes a measurable displacement of the trapped bead, while still localizing it within the optical trap. We present below the key results of our active microrheology experiments.

\vspace{0.3cm}
\noindent \textbf{Strengthening of Laponite networks with increasing Laponite concentration at a fixed aging time:}
Figure~\ref{Concentration} shows the elastic modulus, G$^\prime$, and the viscous modulus, G$^{\prime\prime}$, as a function of stage oscillation frequency $\omega$ at a predetermined aging time $t_w$ = 90 minutes for various Laponite suspension concentrations $C_L$. All the active microrheology measurements are performed using an optically trapped probe particle of diameter $D$ = 1 $\mu$m. At small $\omega$, we note that the $G^{\prime\prime}$ dominates over $G^\prime$ for all Laponite concentrations $C_L$, indicating viscoelastic liquid-like response of the suspension at small frequencies even 90 minutes after preparation. In comparison to our microrheological measurements, representative bulk frequency response data for a Laponite suspension of concentration 3\% w/v at $t_w$ = 90 minutes (Supplementary Fig. S10) shows solid like behavior over the entire frequency range explored. It is well-known that structured soft materials display scale-dependent heterogeneous rheology at comparatively smaller probing length scales that cannot be accessed by bulk rheological approaches~\cite{waigh2016advances}. Even though the material is macroscopically a gel, the trapped probe particle experiences increasingly restricted motion in these fluid regions as the suspension microstructure evolves. Due to the fractal nature of the percolating Laponite network, the sizes of these fluid regions vary. We note here that the microrheological data emphasises the liquid like response of the medium as the immediate neighborhood of the probe particle comprises mostly the water medium. The heterogeneity in the local surroundings also results in heterogeneity in the attachment of the probe particle with the Laponite strands. These factors can account for the discrepancies in the moduli values estimated for the rheological measurements at microscopic and macroscopic length scales.

We see from Fig. 3 that $G^\prime$ increases with increasing $\omega$ and becomes numerically equal to $G^{\prime\prime}$ at a crossover frequency $\omega_{co}$, with a corresponding crossover modulus $G_{co}$. The suspension starts showing viscoelastic solid-like behaviour when the stage oscillation frequency exceeds $\omega_{co}$. We also calculate the frequency-dependent complex viscosities, $\eta^* = \sqrt{(G^{\prime2} + G^{\prime\prime2})} / \omega $, for Laponite suspensions of various $C_L$ at an aging time $t_w$ of 90 minutes (inset of Fig.~\ref{Concentration}). The complex viscosity at low $C_L$ remains almost unchanged with increasing $\omega$, implying that the suspension structure remains invariant under the applied external perturbations. The shear exerted by the probe polystyrene particle is therefore too weak to destructure the underlying Laponite microstructures under these conditions. We note that $\eta^*$ increases with increasing $C_L$, which is in agreement with a previous report on passive microrheology experiments with Laponite suspensions ~\cite{jop2009experimental}. The characteristic relaxation time $\tau$ corresponding to each $C_L$ is obtained from the reciprocal of the crossover frequency $\tau= 2 \pi/\omega_{co}$. Fig.~\ref{G_co}(a) shows that the crossover modulus $G_{co}$ and relaxation time $\tau$ increase monotonically with increasing $C_L$, thereby indicating the strengthening of the underlying Laponite suspension networks with increasing suspension concentration at a fixed Laponite age.

\vspace{0.3cm}
\noindent \textbf{Strengthening of Laponite network structures with increasing suspension age at a fixed suspension concentration:} To understand the temporal evolution of the mechanical moduli of a Laponite suspension of fixed concentration $C_L$ = 2.75\% w/v, we perform active microrheology experiments at various suspension aging times, $t_w$, ranging from 90 to 150 minutes using a probe particle of diameter $D$ = 1 $\mu$m. 
We observe that $G^\prime$ and $G^{\prime\prime}$ show similar frequency responses with varying $t_w$ (raw data plotted in Supplementary Fig. S11) as observed in the earlier experiments in which $C_L$ was varied at a fixed $t_w$ (Fig.~\ref{Concentration}). The crossover modulus $G_{co}$ and characteristic relaxation time $\tau$ for Laponite suspensions at different $t_w$ are shown in Fig.~\ref{G_co}(b). $G_{co}$ increases with aging time $t_w$ and suggests the strengthening of the suspension networks with time due to enhancement in the inter-particle electrostatic interactions. Interestingly, we observe that $\tau$ does not change significantly over the range of $t_w$ explored here. As the probe particle is attached to the Laponite networks, the relaxation process  during stage oscillation depends on both the probe particle size and morphology of the Laponite networks. The interaction between the probe particle connected to the network and the solvent is important for understanding the relaxation process observed in experiments. The approximately constant values of $\tau$ in Fig.4(b) indicate minimal changes in the suspension dynamics within the explored $t_w$ range. We note that all the measurements are performed with Laponite suspensions of ages $t_w \geq $ 90 minutes and Laponite microstructures are presumably already well-formed within the suspension. Therefore, even though the networks continue to strengthen with age, the data in Fig.4(b) indicates that dynamical changes of the kinetically arrested medium are minimal at the microscopic scale.

\begin{figure*}[ht]
\includegraphics[scale=.086]{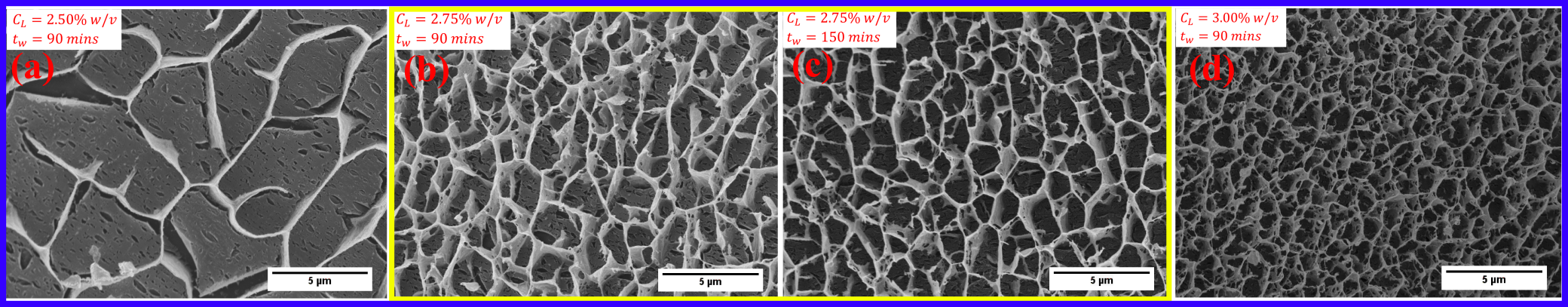} 
\caption{\label{fesem} Cryo-FESEM images of Laponite suspensions of different concentrations $C_L$ and aging times $t_w$ \textbf{(a)} $C_L$ = 2.5 \% w/v, $t_w$ = 90 mins, \textbf{(b)} $C_L$ = 2.75 \% w/v, $t_w$ = 90 mins, \textbf{(c)} $C_L$ = 2.75 \% w/v, $t_w$ = 150 mins, \textbf{(d)} $C_L$ = 3.0 \% w/v, $t_w$ = 90 mins.}
\end{figure*}

\begin{figure}[ht]
\includegraphics[scale=0.7]{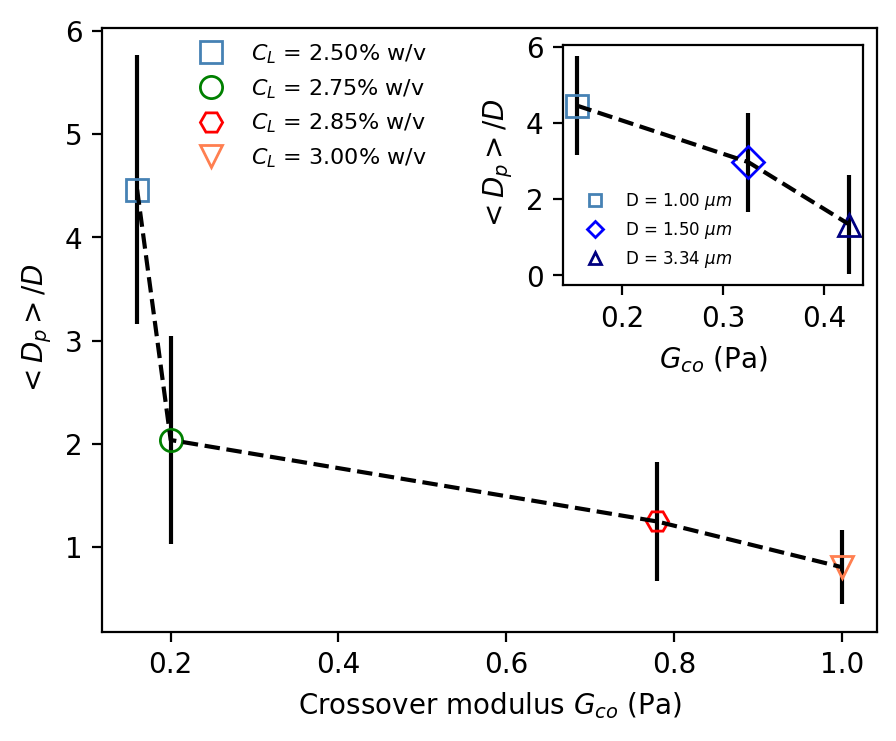} 
\caption{\label{Dp} Pore size ratio $<D_p>/D$, the ratio of average pore diameter $<D_p>$ and the diameter of the trapped particle ($D$=1 $\mu m$), is plotted vs. crossover modulus $G_{co}$ obtained from OT-based active microrheology for different Laponite concentrations $C_L$ at $t_w$ = 90 min. In the inset, the pore size ratio is plotted for different sizes of probe particles D = 1 $\mu$m, 1.5 $\mu$m, 3.34 $\mu$m for Laponite suspension concentration $C_L$ = 2.5\% and age $t_w$ = 90 min. }
\end{figure}

\vspace{0.3cm}
\noindent \textbf{Probe size dependence of measured Laponite viscoelasticity:}
We next explore the lengthscale dependent microrheology of Laponite suspensions by varying the sizes $D$ of the trapped colloidal probe particles. Active microrheology measurements are performed to study Laponite suspensions of  $C_L$ = 2.5\% w/v and $t_w$ = 90 minutes while trapping probe particles of sizes 1, 1.5 and 3.34 $\mu$m. The variations of $G^{\prime}$ and $G^{\prime\prime}$ with stage oscillation frequencies show similar trends (Supplementary Figure S12) as observed earlier (Fig.~\ref{Concentration} and Supplementary Figure S11). The elastic modulus $G^{\prime}$ and viscosity $\eta^*$ of the underlying Laponite suspension measured using a larger probe is higher when compared to that measured with smaller probes. Since the surface area of a bigger bead is larger, the connectivity between the bead and suspension microstructure is stronger. The drag on a probe particle therefore increases as its size becomes larger. We note that active microrheology with bigger probes results in estimates of longer characteristic relaxation times $\tau$ and higher crossover moduli $G_{co}$ of the underlying Laponite suspension medium, as displayed in Fig.~\ref{G_co}(c). Therefore, active microrheology experiments can reveal the length-scale dependence of mechanical properties which, in turn, are directly dependent on the connectivity between the probe and Laponite particles. Probe size dependent microrheology of Laponite suspensions was previously investigated by Oppong et al. \cite{PhysRevE.78.021405} where they characterised the gelation point of Laponite suspensions using passive micro rheology. They found that the gelation point, which marks the transition point where the sample becomes solid-like after initially exhibiting liquid-like viscoelasticity, decreases with an increase in probe size following a power law with an exponent of -0.15. However given the spatial heterogeneity of the Laponite network, and the heterogeneity in attachment of the probe to the Laponite network strand, extracting such a scaling relation as in \cite{PhysRevE.78.021405} is not possible from our experiments.

\vspace{0.3cm}
\noindent \textbf{Correlating pore sizes of Laponite networks with suspension dynamics:}
We note from Fig. 2(a) and Supplementary Fig. S5 that the sizes of probe particles are comparable with the characteristic structural lengthscales of the suspension. We directly visualize Laponite microstructures in suspensions of different $C_L$ and $t_w$ using cryo-FESEM. Figures~\ref{fesem}(a-b,d) show that microstructures are denser for suspensions with higher $C_L$. In contrast, we see that the network morphology does not change appreciably when suspensions of fixed concentration are imaged at different $t_w$ (Figures~\ref{fesem}(b-c)). We calculate the characteristic lengthscales of the Laponite suspension microstructures by extracting an average pore diameter, $<D_p>$, measured according to the protocol discussed in Supplementary Information Section ST10 for all the network structures imaged using cryo-FESEM. We observe that $<D_p>$ decreases with increasing $C_L$ but remains almost unchanged with increasing $t_w$. This result is in agreement with passive microrheology results reported earlier \cite{zheng2020microstructure, vyas2016passive} and implies a correlation between the sample structure and its mechanical response at microscopic lengthscales.

Figure~\ref{Dp} consolidates the OT and cryo-FESEM data analyses to display the variation of the average dimensionless pore diameters, $<D_p/D>$ where $<D_p>$ is obtained from cryo-FESEM and normalized by the probe size D to yield a dimensionless lengthscale, as a function of the crossover modulus $G_{co}$ extracted from our active microrheology experiments. We note from Fig.6 that the lower values of $G_{co}$ correspond to data analyzed from active  microrheology measurements performed with samples of lower $C_L$. We observe a clear inverse correlation between $<D_p>/D$ and the crossover modulus $G_{co}$. This implies that Laponite suspensions having denser microstructures at larger $C_L$ show increasingly viscoelastic solid-like behavior and have larger characteristic relaxation times. We also show the correlation between dimensionless lengthscale $<D_p>/D$ and characteristic relaxation time $\tau$ in Supplementary Fig. S14. We note from the inset of Fig. 6 that for a fixed Laponite concentration of 2.75\% w/v, $<D_p>/D$ decreases with increasing crossover modulus $G_{co}$ for various probe particle sizes $D$. This observation clearly shows that active microrheology measurements are probe size-dependent and soft glassy Laponite suspensions having finite rigidities exhibit lengthscale dependent mechanical properties.  This observation agrees with a previous report on passive microrheology of Laponite  suspensions~\cite{probeize}. The length-scale dependent mechanical properties of Laponite suspensions is a direct consequence of the heterogeneous network structures that  evolve in aging Laponite suspensions. Interestingly, a theoretical work that proposed an elastic model to describe the hierarchical arrest of attractive colloidal particles demonstrated that it is possible to distinguish between the microscopic (particle-scale) and mesoscopic (cluster-scale) contributions to the macroscopic shear modulus \cite{zaccone2009elasticity}. OT based active microrheology studies can therefore serve as an excellent starting point to quantify such spatiotemporal heterogeneities.

\section{Conclusions}

A Laponite suspension shows physical aging, which is manifested as a continuous evolution of the mechanical properties of the suspension with time \cite{aging, aging2}. While in passive microrheology, accurate measurements are limited by the probe-size dependence of the diffusive dynamics, the application of external forces in active microrheology allows the measurement of mechanical properties that are inaccessible using the former method. In this report, we systematically perform oscillatory active microrheological experiments with aging aqueous Laponite suspensions using an optical tweezer. The aim of our work is to uncover correlations between the morphologies of the aging Laponite suspension structures and the probe particle dynamics at micrometer lengthscales.

Polystyrene particles interact with Laponite particles and attach physically to the suspension microstructures. We visualize this attachment using cryo-FESEM imaging. Typically, the size of the probe used in tweezing experiments should be larger than the characteristic lengthscale of the underlying system \cite{tassieri2016microrheology}. When the probe particle interacts with the network, microrheological responses can be extracted even when the characteristic lengthscales of the suspension microstructures are larger in comparison to the probe size~\cite{VALENTINE20044004}. Controlled sinusoidal oscillations of the sample stage, even as the probe particle is optically trapped and attached to the microstructures of the suspension, allow us to extract the viscoelastic properties of the underlying medium. We note that the measured complex viscosities of the suspension do not change appreciably over the range of frequencies explored in our experiment, particularly for the samples of lower concentrations. The Laponite medium is not therefore destructured by the shear forces exerted by the trapped probe particle. Our microrheology data establishes that the suspension networks can strengthen significantly when either suspension concentration or age is changed. With increasing suspension age and concentration, the Debye layer surrounding each Laponite particle shrinks \cite{saha2015dynamic}, resulting in increased participation of the particles in the branches of the Laponite gel network due to enhancement in the particle edge-face  attractive interactions. While the crossover modulus $G_{co}$ measures the strengthening of the gel network due to enhanced Laponite particle self-assembly, the relaxation time $\tau$ provides an estimate of the dynamics of the probe particle trapped in the network pore. We show here that while the network strengthens considerably when suspension concentration, age and probe particle diameter are increased, the dynamical timescale shows a comparatively modest change over the parameter space explored. Our observations such as enhancement of moduli with increase in suspension concentration and age resemble the results extracted from bulk rheology experiments \cite{bonn2002rheology} but have never been reported at microscopic lengthscales using active microrheology. It has been predicted that the macroscopic rheology of arrested states depends on the ratio of particle cluster size and dimension of an individual particle \cite{zaccone2009elasticity}. Interestingly, this would result in lengthscale dependent rheology as demonstrated here and also previously reported for other soft materials in the literature in different contexts \cite{probeize, liu2006microrheology, kurniawan2010image}.

We systematically demonstrate that the rheological properties of soft glassy Laponite suspensions \cite{mourchid1995phase, tanaka2004nonergodic}, are correlated with the underlying suspension structure, $viz.$, the average network pore sizes of the suspension networks imaged using Cryo-FESEM. We show that over the range of aging times investigated here for a fixed Laponite concentration, the underlying dynamics is approximately invariant. When the Laponite concentration $C_L$ increases, the network structure, visualized using cryo-FESEM, becomes denser. We demonstrate that the crossover modulus of the suspension, obtained from active microrheology measurements, is inversely correlated with the average pore size of the Laponite network. Such a correlation between the lengthscale characterising the suspension network and the characteristic relaxation time of the suspension at micrometer scales is shown for the first time to the best of our knowledge using optical tweezer based active microrheology.

Bulk and micrometer-scale rheology results are expected to agree if probe particles are much larger compared to the characteristic length scale of microstructural features in the suspension \cite{Tassieri2019MicrorheologyWO}. Previous passive microrheology experiments with polymer gels and associating polymer systems observed that the probe particle size had to be about 20 times larger than the characteristic structural length scale (pore size) to obtain continuum viscoelastic moduli values \cite{lu2002probe, schnurr1997determining}. The increasing $G_{co}$ values with probe radius (Fig. 4(c)) in our experiments indicate the existence of a length scale where the measured rheology saturates to macrorheological values. In this regime, the local heterogeneities are expected to be averaged out, resulting in an approximately homogenous medium over the length scales probed. The inconsistency in the connectivity of the Laponite networks with the probes also becomes inconsequential over this length scale. However, using optical tweezer based active microrheology, accessing the intermediate regime where the micrometer-scale estimation meets macro-scale measurements is challenging to probe due to the limited trap stiffness \cite{montange2013optimizing}. 

Nevertheless, our results can be applied to study the effects of local sample heterogeneities during transport of micron-sized particles through complex environments, for example, in drug delivery ~\cite{viseras2008biopolymer}. Laponite is utilised as a rheological modifier  \cite{oleyaei2018novel,davila2017laponite} in material processing applications, and knowledge of micron-scale local dynamics is necessary for achieving the desired outcomes. A possible extension of the present work could involve the setting up of multiple traps to precisely study the short-range suspension-mediated interactions \cite{meiners1999direct,di2008hydrodynamic}, or to understand the effects of quenched defects in a viscoelastic medium \cite{marchetti2002viscoelasticity}. Non-linear vicoelasticity \cite{gomez2014probing} of aging Laponite suspensions can be explored even at microscopic lengthscales by applying large amplitude stage oscillations. These studies could shed additional light on the dynamics of local relaxations and defects in kinetically arrested states of colloidal suspensions. Indeed, further studies are required to produce quantifiable data, including microscopically probing Laponite suspensions of higher aging times and concentrations over a range of probe sizes. Such investigations would necessitate active microrheology studies using different techniques such as magnetic tweezers or atomic force microscopy, which can exert larger forces locally \cite{neuman2008single}. We further note that different additives and external fields can be incorporated in Laponite suspensions to fine-tune the rheological response \cite{misra}. The present work could therefore be extended to estimate micron-scale mechanical responses of suspensions in the presence of a range of additives or due to the application of various external fields.

\noindent \textbf{Author Contributions: Rajkumar Biswas:}  Methodology, Formal analysis, Software, Validation, Investigation, Writing – original draft, Visualization, Data curation. \textbf{Vaibhav Raj Singh Parmar:} Investigation, Software, Writing – original draft, Validation, Visualization, Data curation. \textbf{Anson G. Thambi:}  Investigation, Software, Validation, Writing – original draft, Data curation. \textbf{Ranjini Bandyopadhyay:} Conceptualization, Methodology, Validation, Writing – review \& editing, Supervision, Funding acquisition, Project administration.

\noindent \textbf{Data availability:} The data that support the findings of this study are available from the corresponding author upon reasonable request.

\noindent \textbf{Conflicts of interest:} The authors declare that there are no conflicts of interest to report in this work. 

\noindent \textbf{Acknowledgement:} We would like to thank MD Arsalan Ashraf and Abhishek Ghadai for useful discussions. We want to thank Yatheendran K.M. for help with Cryo-FESEM. We would also like to thank Vasudha K.N for calorimetry measurements.

\bibliographystyle{vancouver}
\bibliography{sorsamp}

\end{document}


\renewcommand{\figurename}{Supplementary Fig.}
	\setcounter{table}{0}
	\renewcommand{\thetable}{S\arabic{table}}%
	\renewcommand{\tablename}{Supplementary Table}
	\setcounter{figure}{0}
	\makeatletter 
	\renewcommand{\figurename}{Supplementary Fig.}
	\setcounter{figure}{0}
	\makeatletter 
	\renewcommand{\thefigure}{S\arabic{figure}}
	\setcounter{section}{0}
	\renewcommand{\thesection}{ST\arabic{section}}
	\setcounter{equation}{0}
	\renewcommand{\theequation}{S\arabic{equation}}
 
	\title{\color{black}\textbf{\underline{Supplementary Information}\\ Correlating microscopic viscoelasticity and structure of an aging colloidal gel using active microrheology and cryogenic scanning electron microscopy}}

	\author[1, $\dagger$]{Rajkumar Biswas}
	\affil[1]{\textit{Soft Condensed Matter Group, Raman Research Institute, C. V. Raman Avenue, Sadashivanagar, Bangalore 560 080, INDIA}}
	\author[1, $\ddagger$]{Vaibhav Raj Singh Parmar}
        \author[1, $\ddag$]{Anson G Thambi}
	\author[1,*]{Ranjini Bandyopadhyay}
	\date{\today}
	
	\footnotetext[2]{rajkumar@rri.res.in}
	\footnotetext[3]{vaibhav@rri.res.in}
        \footnotetext[4]{anson@rrimail.rri.res.in}
	\footnotetext[1]{Corresponding Author: Ranjini Bandyopadhyay; Email: ranjini@rri.res.in}
	\maketitle
	\pagebreak
 
 \section{Schematic diagram of microchannel used in OT experiments}
            \begin{figure}[H]
	 	\includegraphics[width= 5.0in ]{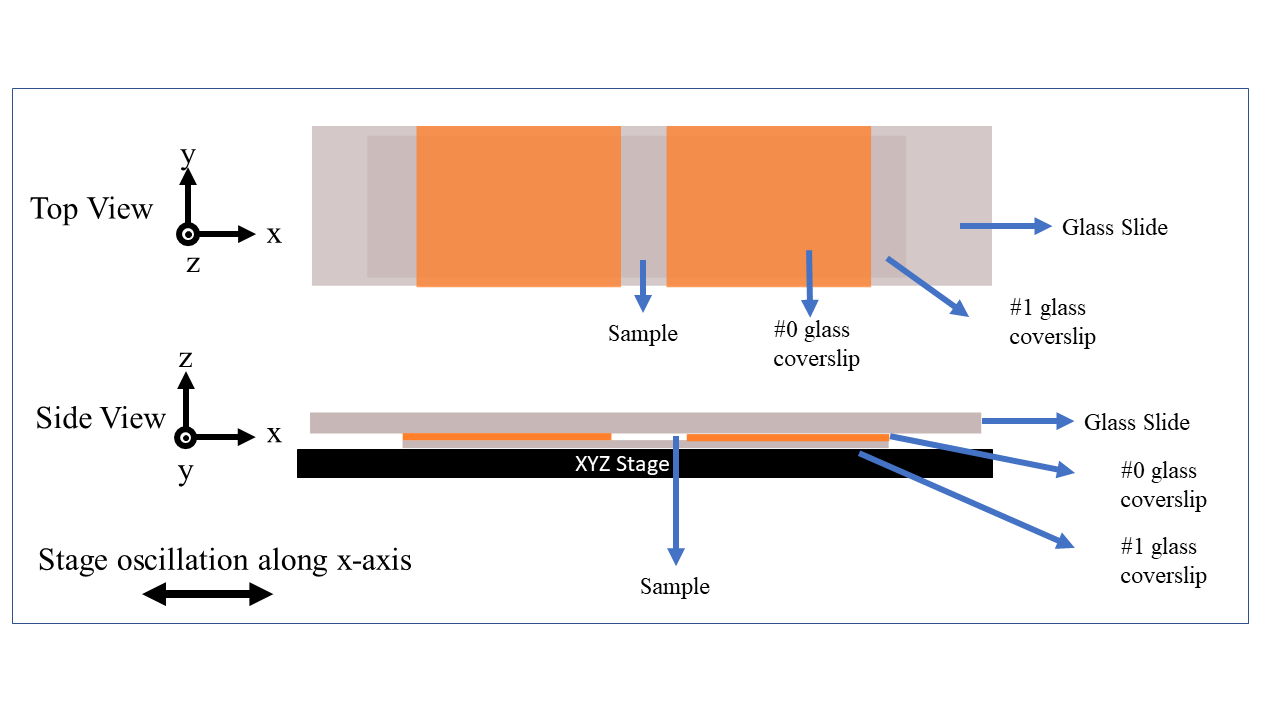}
	 	\centering
	 	\caption{\label{glasschannel} Schematic illustration of the glass microchannel used for the OT experiments. Active microrheology measurements are performed by oscillating the XYZ stage along the x axis.}
	       \end{figure}

            
            \section{Piezoelectric stage calibration: calculation of the voltage-displacement conversion factor for controlled movement of the sample stage}
            \begin{figure}[H]
	 	\includegraphics[width= 5.0in ]{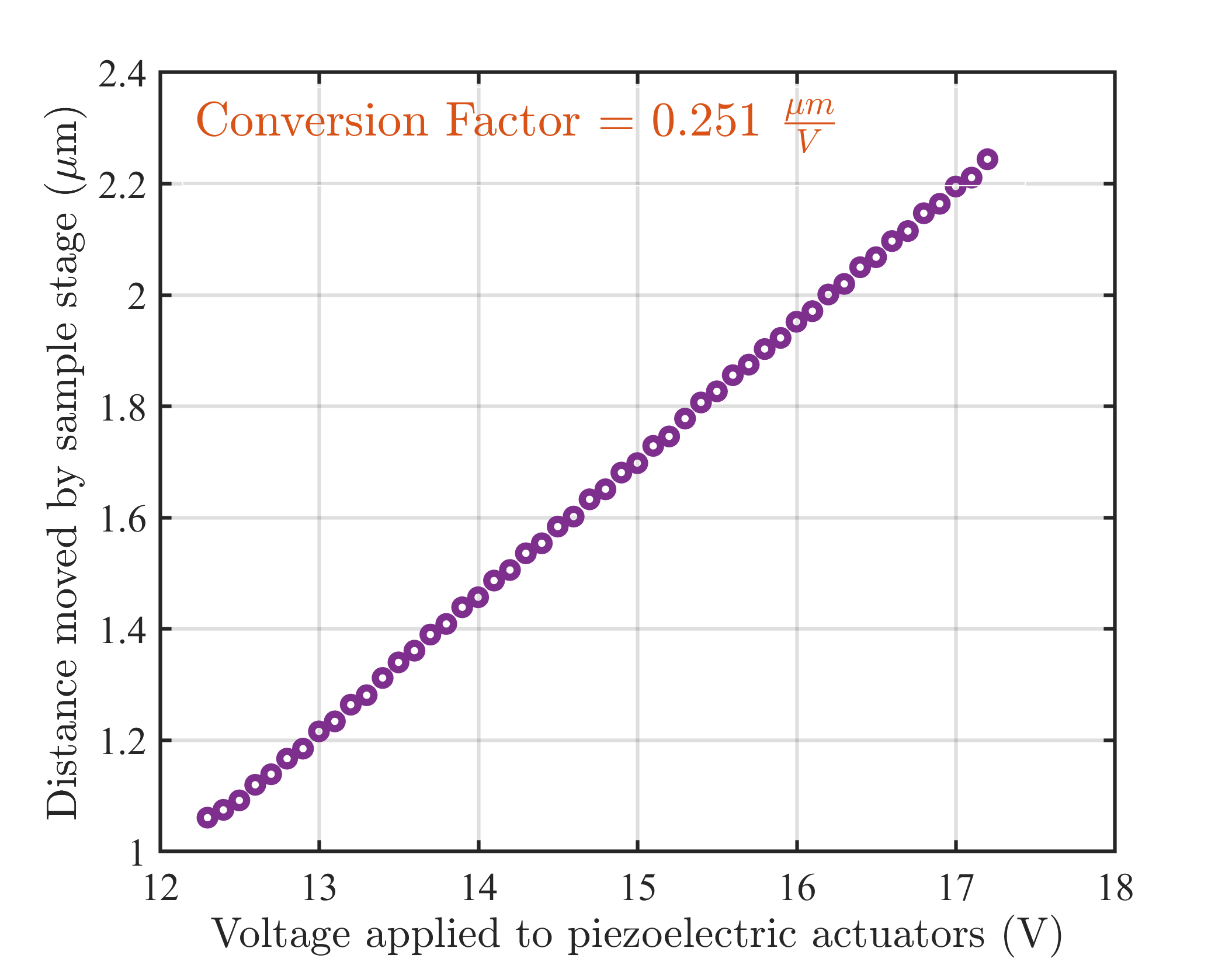}
	 	\centering
	 	\caption{\label{piezo} Distance traveled by the sample stage is measured as a function of the voltage applied to the piezo actuator. The calibration factor, computed from the slope of the curve, is 0.251 $\mathrm{\mu}$m/V.}
	       \end{figure}
	 
            The sample stage consists of a slide holder with a 3-axis XYZ translational stage, piezoelectric actuators and strain gauges. Application of a voltage to the piezoelectric actuators results in the desired oscillation of the sample stage. The strain gauges measure the accurate position of the sample stage feedback for controlled spatial movement. \\
            The conversion factor for calibrating the voltage applied to the piezoelectric actuators to the net distance travelled by the sample stage is next computed.  We prepare a glass microchannel containing a tiny amount of polystyrene beads suspended in 0.5 mM NaCl aqueous solution. The beads get attached to the glass coverslip in a high salt environment which allows us to acquire the bead position using a CMOS camera as we apply a voltage to the piezoelectric actuators. The slope of the plot of stage displacement versus voltage applied to the piezoelectric actuators gives a voltage to displacement conversion factor of 0.251 $\mu$m/V (Supplementary Fig. ~\ref{piezo}). 
            
            \section{Quadrant photodiode (QPD) position calibration}
            \begin{figure}[H]
	 	\includegraphics[width= 6.0in ]{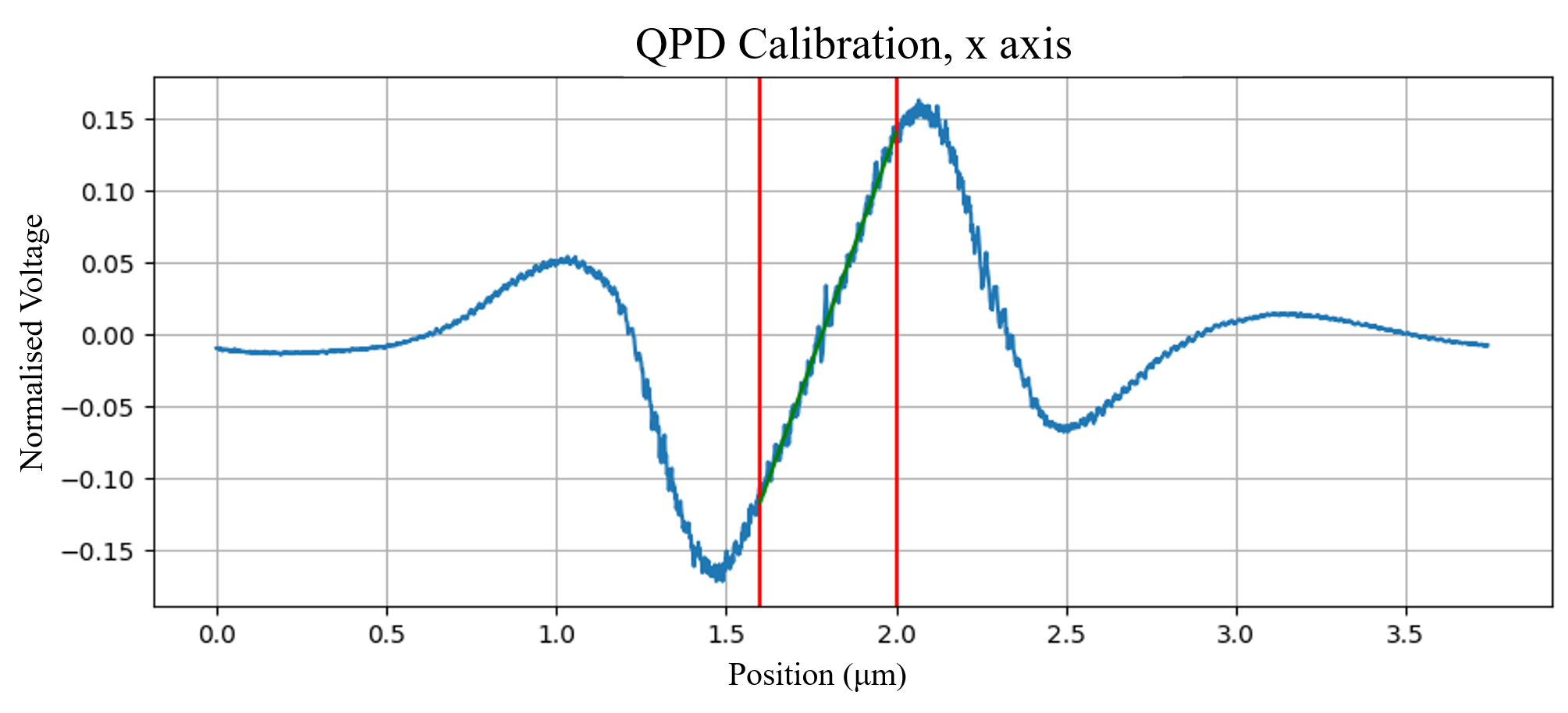}
	 	\centering
	 	\caption{\label{QPD}  Position calibration of QPD is performed by shifting a bead attached to a glass coverslip along the x-axis by a predetermined distance from the laser's beam center. The QPD signal is acquired during this movement and plotted against the motion of the trapped bead. We determine the sensitivity of the QPD, 1.51 nm/QPD unit, by fitting the linear region of the curve with a straight line.}
	       \end{figure}
        
                High-frequency position measurement of an optically trapped bead using a quadrant photodiode requires a relation between QPD output voltage and bead position data. To determine this relation, also called QPD sensitivity, we moved a bead attached to the glass coverslip along the x and y axes such that the bead passes through the center of the laser spot while QPD output voltages are simultaneosuly measured. The QPD output voltage is linearly related to the bead position in a narrow trap region between the two parallel red lines as displayed in Supplementary Fig.~\ref{QPD}. The slope of the linear region yields a QPD sensitivity of 1.51 nm/QPD unit.

                \section{Measurement of trap stiffness using power spectrum method}
                \begin{figure}[H]
	 	 \includegraphics[width= 6.0in ]{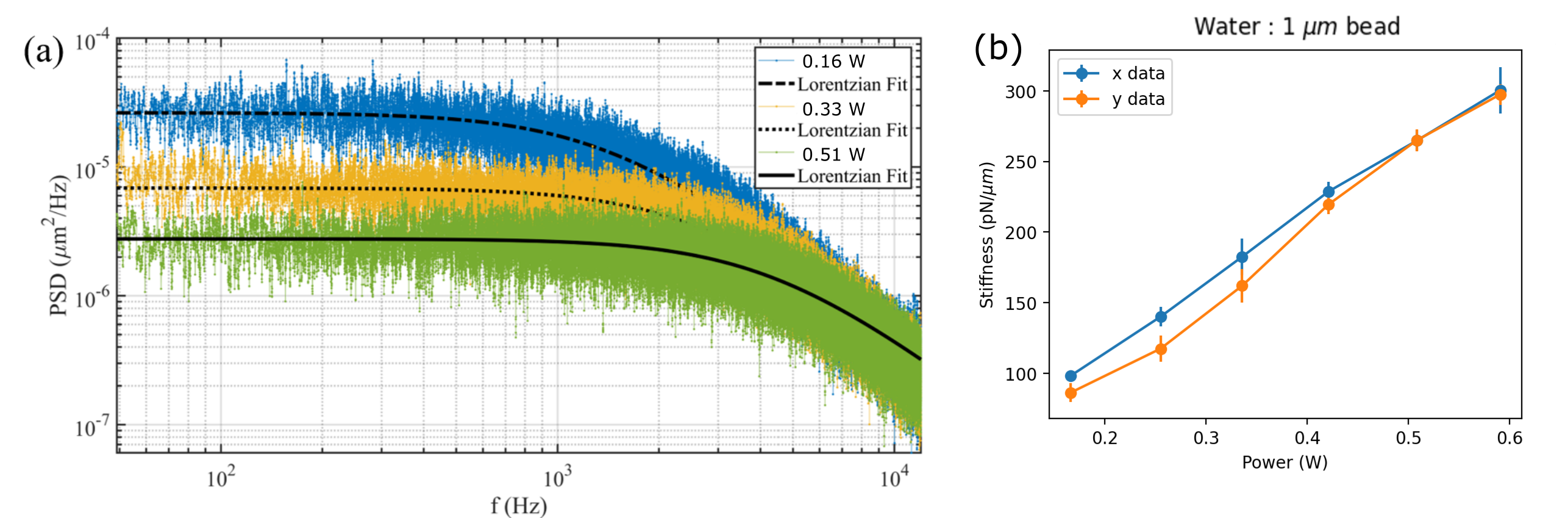}
	 	 \centering
	 	 \caption{\label{psd} (a) Power spectral density (PSD) is plotted for three values of incident laser power and fitted with Lorentzian functions to determine the trap stiffness. (b) Trap stiffness is plotted as a function of laser power for displacements along x and y directions.}
	       \end{figure}
The trap stiffness depends on incident laser power~\cite{calibraton}. We trap a polystyrene bead (1 $\mathrm{\mu}$m) suspended in water and measure its movement at a sampling frequency of 50 KHz. The Fourier transform of the bead position as shown in Supplementary Fig.~\ref{psd}(a) is fitted to the Lorentzian function, $PSD = A/(f^2 + {f_c}^2)$, where $A$ depends on the surrounding temperature, $f$ is the frequency, $f_c = \kappa/2 \pi \gamma$ is the corner frequency and $\gamma$ is the coefficient of the viscosity of the medium ~\cite{calibraton}. The trap stiffness $\kappa$ increases with laser power as shown in Supplementary Fig.~\ref{psd}(b).

\pagebreak

\section{Cryo-FESEM}
    Cryo-FESEM images of PS particles of different sizes attached to Laponite networks. 
    \begin{figure}[H]
	 	 \includegraphics[width= 6.0in ]{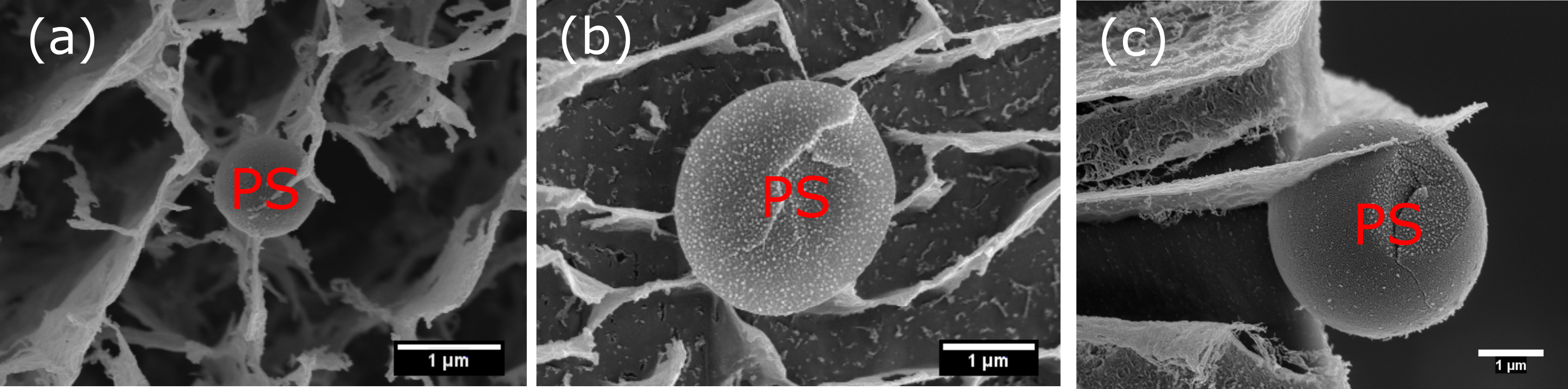}
	 	 \centering
	 	 \caption{\label{itc} Polystyrene (PS) particles attached to the branches of the Laponite suspension networks (white connected regions) for particles of diameter (a) 1 $\mu$m (b) 1.5 $\mu$m and (c) 3.34 $\mu$m.}
	       \end{figure}

\begin{figure}[h]
	 	 \includegraphics[width= 5.0in ]{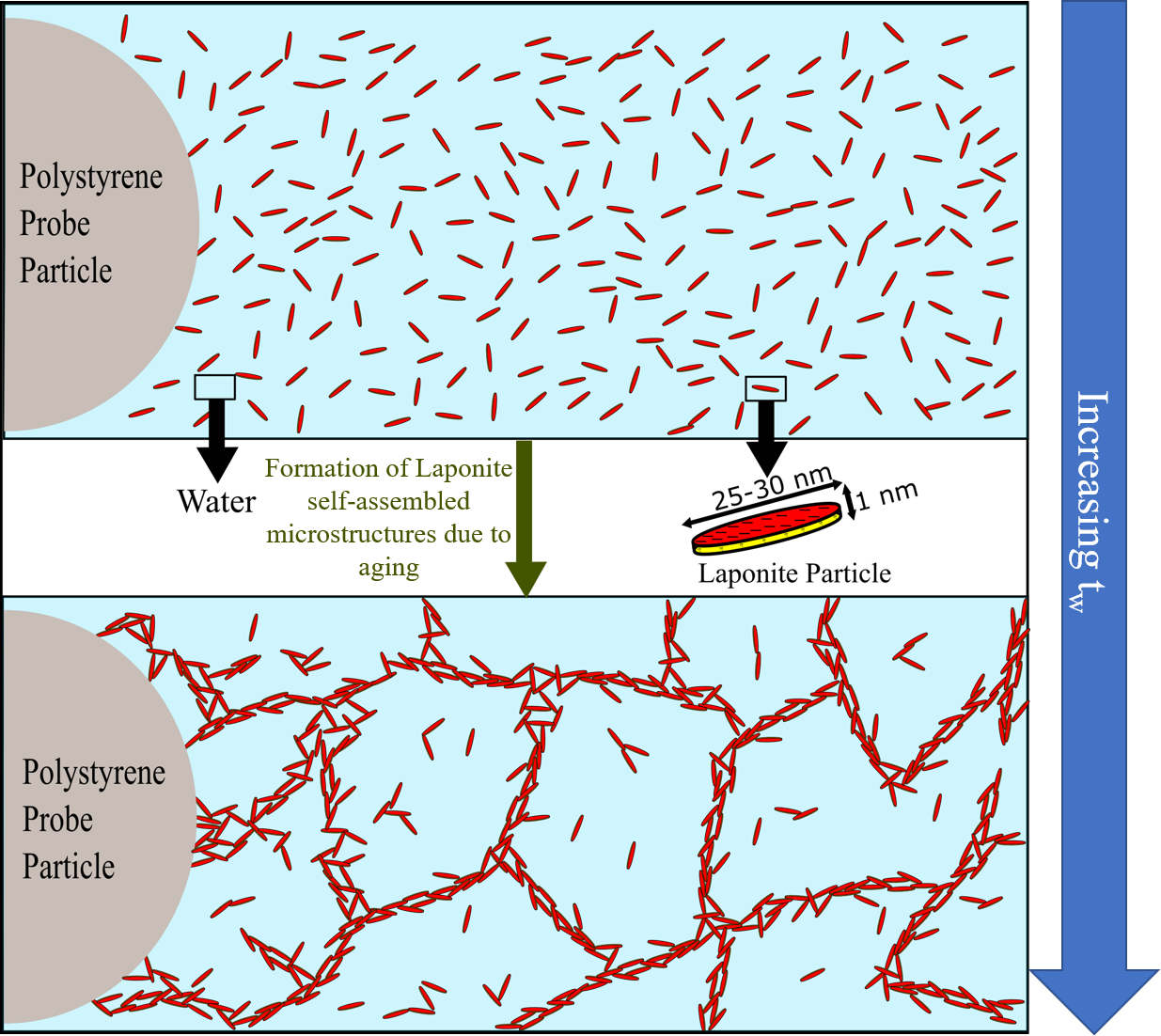}
	 	 \centering
	 	 \caption{\label{microrheology} Schematic illustration showing the temporal evolution of Laponite microstructure in the presence of a polystyrene probe particle. Top panel: The probe particle attaches to the Laponite microstructure, with the number of contacts depending on particle size. Bottom panel: Individual Laponite particles self-assemble at higher aging times to form network structures through house of cards and overlapping coin assemblies.}
   \end{figure}

        \begin{figure}[H]
	 	 \includegraphics[width= 6.0in ]{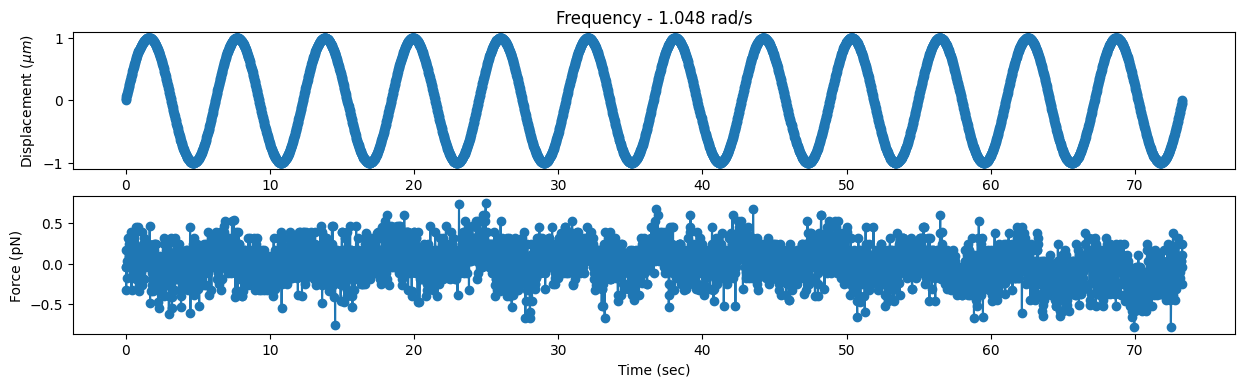}
	 	 \centering
	 	 \caption{\label{NoForce} Top: Low frequency (= 1.048 rad/sec) stage oscillation of amplitude 1 $\mathrm{\mu}$m and Bottom: the corresponding force experienced by the trapped bead of diameter 1 $\mathrm{\mu}$m. The resultant force is not oscillatory and cannot be analyzed to determine the mechanical properties of the Laponite suspension at this frequency.}
   \end{figure}

    \section{Error calculation}

For each measurement, 12 sinusoidal oscillations have been implemented and the force on the trapped bead has been measured. We have excluded the initial and the final oscillations to avoid any artefact driven by transient signals. To compute error bars in measuring $G^\prime$ and $G^{\prime \prime}$, we have therefore used the ten intermediate oscillations. The computed values of $G^\prime$ and $G^{\prime \prime}$ are seen to be reasonably consistent and the deviations for these ten different cycles are presented as error bars.

    \begin{figure}[H]
	 	 \includegraphics[width= 6.0in ]{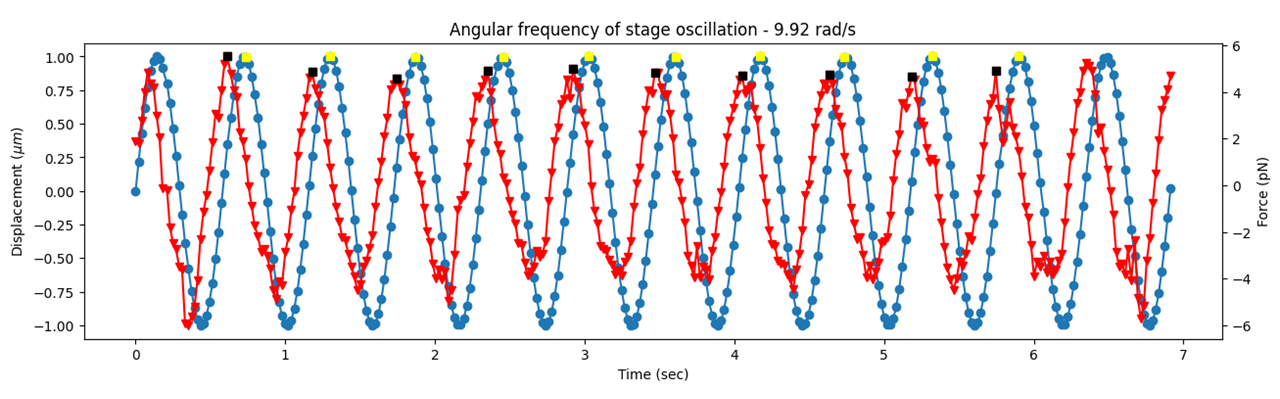}
	 	 \centering
	 	 \caption{\label{error1} The blue curve with solid circles indicates the stage position with time. The peaks in the signal are detected using scipy module in Python. The peaks in the stage oscillation signals are indicated using yellow squares. The force on the trapped particle is shown by the red curve with red triangles. The black squares indicate the peaks of the force signal.}
   \end{figure}

Ten peaks, after excluding the first and last oscillations, are used to calculate the time difference between the peaks of the applied stage displacement and the corresponding force measurement (Supplementary Fig.S8). This results in 10 estimates of time differences which are used to compute the mean and standard deviation in the measured phase angle using the protocol below,

$$t_{diff} = mean (t_1, t_2, …. ,t_{10})$$
$$t_{error}= standard\:deviation (t_1,t_2, …. ,t_{10})$$

Phase angles in degree are estimated computing,

$$\phi_{diff}=360^\circ * f * t_{diff} ,\: \phi_{error}=360^\circ * f * t_{error} $$

where $f$ = stage oscillation frequency (in Hz)

The mean and standard deviation of the phase angles calculated according to the method as described above are used to calculate the mean and standard deviation values of the elastic and viscous moduli using the relations (Eqns. 1 and 2 of the manuscript),

$$ G^{\prime} = \frac{F_{max}}{6 \pi RX_{max}} cos(\frac{\pi \phi_{diff}}{180}), \: \Delta G^{\prime} = \frac{F_{max}}{6 \pi RX_{max}}sin(\frac{\pi \phi_{diff}}{180})(\pi \phi_{error}/180)$$

$$ G^{\prime \prime} = \frac{F_{max}}{6 \pi RX_{max}} sin(\frac{\pi \phi_{diff}}{180}), \: \Delta G^{\prime \prime} = \frac{F_{max}}{6 \pi RX_{max}}cos(\frac{\pi \phi_{diff}}{180})(\pi \phi_{error}/180)$$

where $\Delta \phi$ = $\frac{\pi \phi_{diff}}{180}$.

$G^{\prime}$ and $G^{\prime \prime}$ with error bars are plotted in transparent colors in the Supplementary Fig. S9. The maximum error propagation while computing $G_{co}$ and $\omega_{co}$ is estimated by considering the maximum ranges of errors of these two variables, marked using the dashed lines for both horizontal and vertical directions. To calculate the errors in crossover modulus $G_{co}$ and relaxation time $\tau_{co}$, the following relations have been used:

$$G_{co} = (G_{co2} + G_{co1})/2, \: \Delta G_{co} = (G_{co2} - G_{co1})/2$$
$$\omega_{co} = (\omega_{co2} + \omega_{co1})/2, \: \Delta \omega_{co} = (\omega_{co2} - \omega_{co1})/2$$
$$ \tau_{co} = 2\pi/\omega_{co}, \: \Delta \tau_{co} = (2\pi/\omega_{co}^2) \Delta \omega_{co}$$

    \begin{figure}[H]
	 	 \includegraphics[width= 5.0in ]{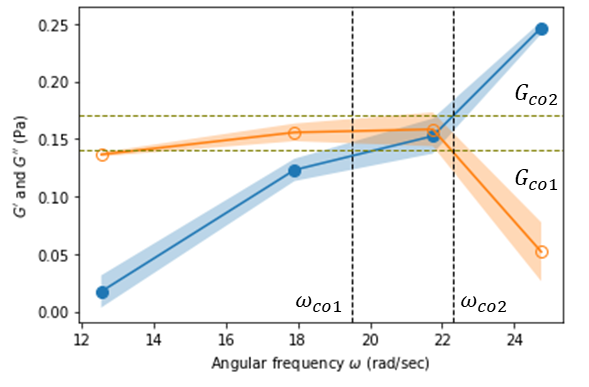}
	 	 \centering
	 	 \caption{\label{error1} $G^\prime$ (solid blue symbols) and $G^{\prime \prime}$ (hollow orange symbols) and the corresponding error bars are shown by the shaded regions of the same colour. The maximum possible ranges of the crossover values of the moduli and angular frequencies are marked using the dashed horizontal and vertical lines.}
   \end{figure}

    \section{Bulk rheology measurement of Laponite suspension}
        \begin{figure}[H]
	 	 \includegraphics[width= 4.0in ]{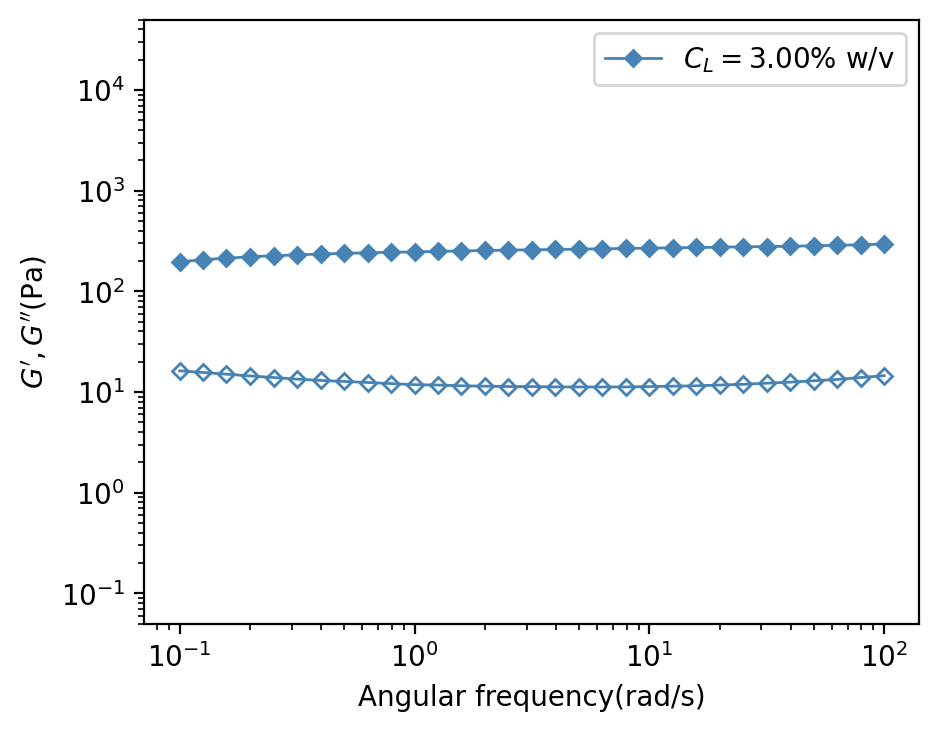}
	 	 \centering
	 	 \caption{\label{bulk} Macroscopic or bulk elastic ($G^\prime$, solid symbols) and viscous ($G^{\prime \prime}$, hollow symbols) moduli of a Laponite suspension of concentration $C_L$ = 3.00\% w/v at aging time $t_W$ = 90 min as measured using a rheometer.}
   \end{figure}

        \section{Active microrheology measurements using 1 $\mathrm{\mu}$m probe particle for Laponite suspensions of concentration $C_L$ = 2.75\% w/v at various aging times $t_w$}
        \begin{figure}[H]
	 	 \includegraphics[width= 6.0in ]{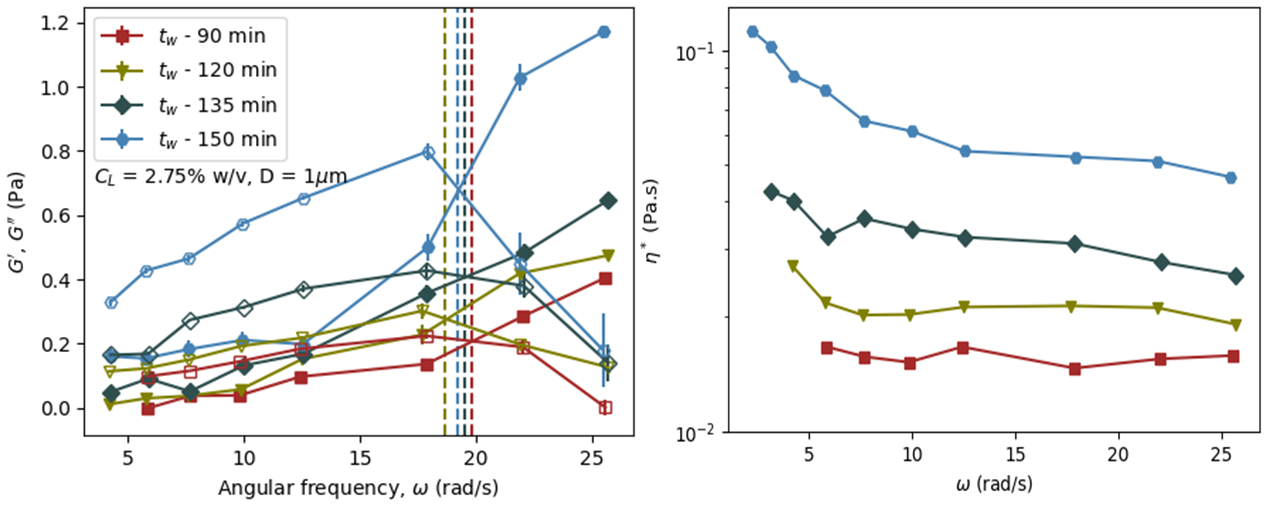}
	 	 \centering
	 	 \caption{\label{VarAgingTime} (a) Elastic and viscous moduli, G$^{\prime}$ (solid symbols) and G$^{\prime \prime}$ (hollow symbols) respectively, are estimated using oscillatory active microrheology for a 2.75\% w/v Laponite suspension at different aging times $t_w$. (b) The values of complex viscosity $\eta^*$, estimated using $\eta^* = \sqrt{({G^\prime}^2+ {G^{\prime\prime}}^2)}/\omega$, are plotted $vs.$ angular frequency, $\omega$, for suspensions of the same aging times as in (a).}
   \end{figure}

        \section{Active microrheology measurements for 2.5\% w/v Laponite suspensions at aging time $t_w$ = 90 minutes using probe particles of different sizes $D$}
        \begin{figure}[H]
	 	 \includegraphics[width= 6.0in ]{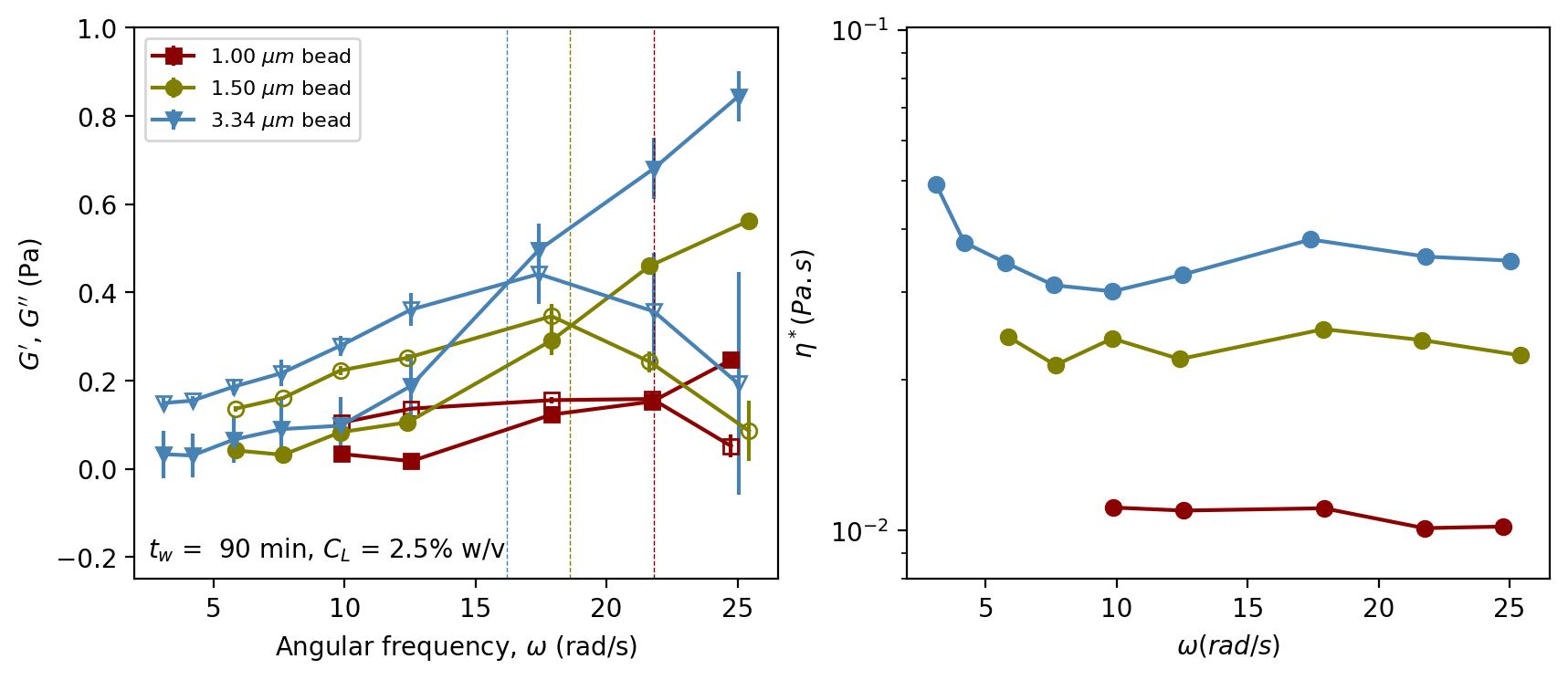}
	 	 \centering
	 	 \caption{\label{varProbeSize} (a) G$^{\prime}$ (solid symbols) and G$^{\prime \prime}$ (hollow symbols) are estimated for a 2.5\% w/v Laponite suspension at aging time $t_w$ = 90 minutes using oscillatory active microrheology with probe particles of different sizes $D$. (b) The values of complex viscosity $\eta^*$ are plotted $vs.$ angular frequency, $\omega$, obtained by analyzing the same experiments.}
   \end{figure}

        \section{Pore size estimation from cryo-FESEM images}

\begin{figure}[H]
	 	 \includegraphics[width= 6.0in ]{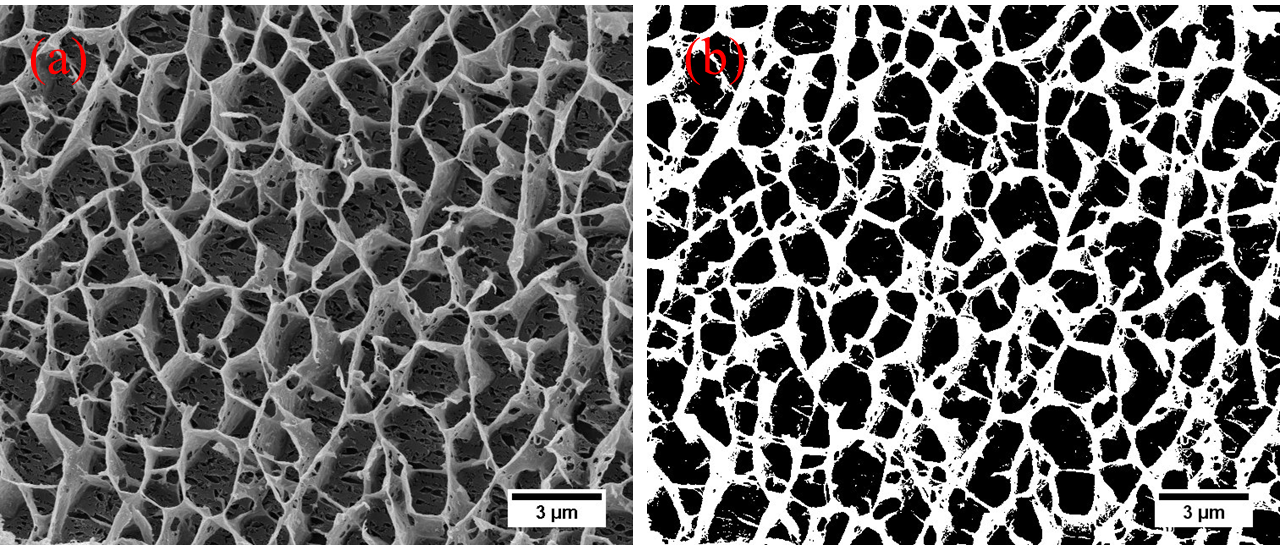}
	 	 \centering
	 	 \caption{\label{imagej} (a)  Cryo-FESEM image of Laponite networks of concentration 2.75\% w/v at 90 min of aging time after converting to gray-scale. (b) Binarized version of image in (a) used to detect network boundaries.}
   \end{figure}

        The acquired cryo-FESEM images (a representative image is displayed in Fig.~\ref{imagej} (a)) are initially binarized (binary image is displayed in Fig.~\ref{imagej}(b)) to emphasise the Laponite networks using an inbuilt module in the ImageJ software.  The area within each pore is measured using ImageJ. The equivalent pore diameter is then estimated by assuming that the measured pore is circular. The average diameter $<D_p>$ is determined by averaging the diameters of various pores (between  $\approx$ 10-50). A dimensionless characteristic lengthscale $<D_p/D>$ is estimated by dividing by the probe particle size. The cryo-FESEM image and the corresponding binary image are shown below in Fig. S13.
        
\section{Correlation between microstructure and relaxation time}

    \begin{figure}[H]
	 	 \includegraphics[width= 4.0in ]{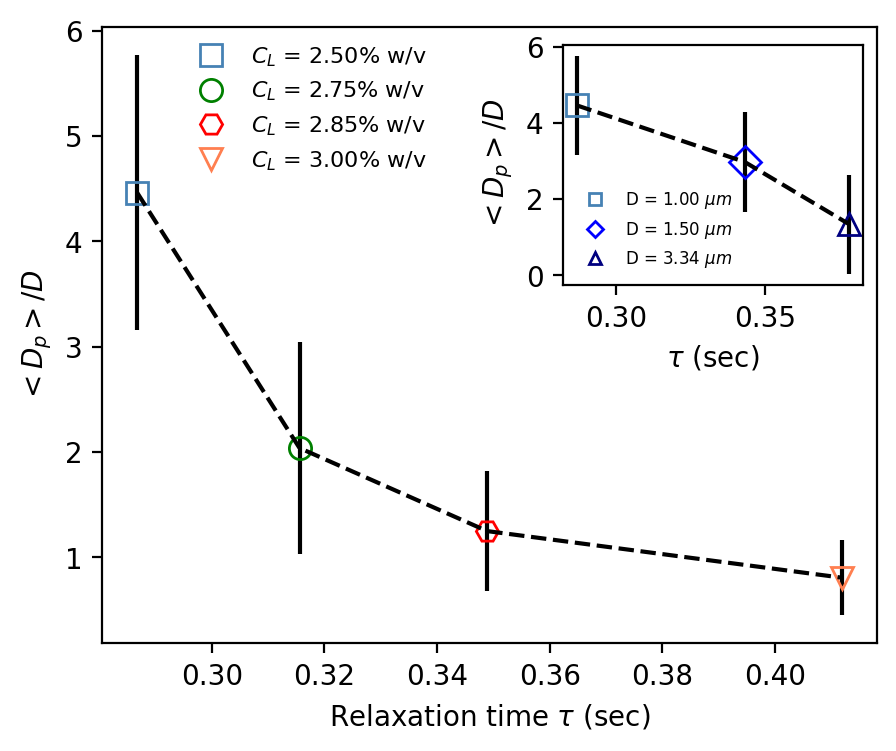}
	 	 \centering
	 	 \caption{\label{last} Pore size ratio $<D_p>/D$, the ratio of average pore diameter $<D_p>$ and the diameter of the trapped particle ($D$=1 $\mu m$), is plotted vs. characteristic relaxation time $\tau$.}
	       \end{figure}

\bibliographystyle{unsrt}
\bibliography{sorsamp}        
	
    \vspace{1 cm}